\documentclass{JHEP3}

\newcommand{\be}{\begin{equation}}
\newcommand{\ee}{\end{equation}}

\title{More about chiral symmetry restoration at finite temperature in the planar limit.}
\author{R. Narayanan
\\Department of Physics, Florida International University, Miami,
FL 33199, USA\\E-mail: \email{rajamani.narayanan@fiu.edu}}
\author{ H. Neuberger
\\ Rutgers University, Department of Physics and Astronomy,
Piscataway, NJ 08855, USA\\E-mail: \email
{neuberg@physics.rutgers.edu} }

\abstract {In the planar limit, in the deconfined phase, the Euclidean
Dirac operator has a spectral gap around zero. We show that functions
of eigenvalues close to the spectral edge, which are independent
of common rescalings and shifts gauge configuration by gauge configuration,
have distributions described by a Gaussian Hermitian matrix model. 
However, combinations of eigenvalues that are scale and shift invariant only 
on the average, do not match this matrix model. 
}
\keywords{1/N Expansion, Lattice Gauge Field Theories}

\preprint{}

\begin{document}

\section{Introduction.}

At infinite number of colors ($N_c$), $SU(N_c)$ gauge theory undergoes
a first order deconfinement transition at a temperature $T_d$~\cite{pis}.  
If present in the Lagrangian, chiral symmetry is 
spontaneously broken for temperatures $T<T_d$, 
but gets restored for $T>T_d$.
This is reflected by the spectrum of the Euclidean Dirac
operator opening a finite gap around zero, as the temperature decreases
though $T_d$~\cite{prevplb}.

Many consequences of the underlying chiral symmetry have
little or nothing to do with the ultraviolet structure of the gauge
theory. Nowadays~\cite{overlap}, one can study these properties
directly on the lattice, without taking the lattice spacing to zero and
separate the following two questions: a) Does a given qualitative feature
hold on the lattice and, if it does, what are the values of the related 
parameters on the lattice?
b) Does the lattice property survive the continuum limit and, if it does, what
are then the physical values of the relevant parameters ?
We shall only concern ourselves with a
question of type a), regarding the statistical properties of the
spectrum of the Euclidean Dirac operator in the deconfined phase, 
where chiral symmetry is restored and when the number of colors $N_c$
is made very large. As the issue remains somewhat unsettled even on
the lattice, we do not proceed to a related study of type b).

We work on a hypercubic lattice of shape $L_4 L^3$. The gauge action
is of single plaquette type. We identify the transition as described
in our previous work~\cite{prevplb} and use identical notation.
Based on ~\cite{overlap} and ~\cite{ext}, and on the fact that fermion
loops can be ignored at large $N_c$, we have at our disposal, for
each gauge configuration $C$, a lattice Euclidean Dirac 
operator $A$, which is antihermitian, anticommutes with $\gamma_5$, and
whose spectrum is unbounded. 
We shall numerically extract the eigenvalues of 
$A$ that are closest to zero. 
Let
\be
\pm i \lambda_j, ~~j=1,2,....
\ee
be the eigenvalues of $A$ at some fixed $N_c$, gauge coupling,
and $L,L_4$, with
\be
0<\lambda_1<\lambda_2<\lambda_3,....
\ee

At low temperatures chiral symmetry is 
spontaneously broken and in this case the spectrum of $-A^2$
reaches zero. Because of the infinite
number of colors and the lack of relevance of the size of the
system due to large $N_c$ reduction~\cite{nna}, one can think of the
Euclidean Dirac operator ($D$) as a large random anti-hermitian matrix,
whose structure is restricted only by chiral symmetry.
\be
\label{rmtclass}
D=\pmatrix {0 & C\cr -C^\dagger & 0 } \ee In the spirit of
Wigner's approach to complex nuclei, one is lead to write down the
simplest probability distribution for the matrix $C$~\cite{sv},
whose linear dimension is proportional to $N_c$:
\be
P(C)\propto e^{-\kappa \dim (C) Tr C^\dagger C }\label{prob}
\ee
The spectrum of $D$ is of the form $\pm i\xi_i$ with
\be
0 < \xi_1 < \xi_2 < \xi_3, \cdots
\ee
This model will be referred to as chRMM in this paper, where the acronym RMM 
stands for ``random matrix model''. The chRMM describes the spectrum
of $A$ close to zero. 
Chiral symmetry breaking is a direct consequence, 
giving it the
appearance of a generic phenomenon. 

In the deconfined regime there is a gap around zero in the spectrum of $A$
and chiral symmetry is restored. One might have thought
that the opening of a gap can be incorporated into an extended random
matrix model~\cite{step}. However, for $T<T_d$ random matrix
theory applies also at finite $N_c$, as a result of Effective Chiral
Lagrangian considerations. This argument does not extend to high
temperatures ~\cite{urs}, where there is no energy regime dominated by
Goldstone particles~\cite{rmt}.

The edge of the gap is ``soft'' in that $\lambda_1$ fluctuates into the gap
region without constraint. The universal features of the spectrum might then
be as well described by the edge of the spectrum of a random hermitian
matrix, $H$, which is not necessarily of the form 
$\sqrt{C^\dagger C}$ (eq. (\ref{rmtclass})) with
a gaussian probability distribution for the complex matrix $C$. Rather, we can
take $H$ itself to be Gaussianly distributed, yielding the most basic of
all RMM-s~\cite{sosh}. The upper edge of the spectrum of this
model ($h_0$) plays no role in the following. 
We denote the highest ordered eigenvalues of $H$ as follows:
\be
...\xi_3 <\xi_2 <\xi_1 < h_0
\ee

We shall consider up to
six eigenvalues
in the RMM and lattice data. 
In order to obtain results in the RMM, we use a 
$100\times 100$ matrix and generate $10,000$ configurations.
In principle one could derive the required information analytically within
the RMM, but we doubt that this is necessary or even useful at this point. 
The lattice data consists from sets of few hundred to order one thousand
samples, and this limits the precision at which a match between data and RMM
can at all be expected.

Much of the lattice data obtained in this paper is for
$b=0.36$, $N_c=47$, $L=6$ and $L_4=4$, where we have
stored up to six eigenvalues per gauge configuration.
From our previous work~\cite{prevplb},
we know that at large $N_c$ the lattice system 
is in the deconfined phase for these values of $b$ and $L,L_4$.
$b$ is the inverse 't Hooft parameter. We also know that
large $N_c$ reduction holds in the sense that the large $N_c$ limit
is independent of $L$ for $L\ge 6$.

As $N_c$ becomes very large, we are left with an open question for
$T>T_d$: is the spectrum of the Euclidean Dirac operator described by
some RMM ? Previous work has shown that the correlation between level
energy fluctuations is incompatible with a {\it simple}
RMM~\cite{prevplb}.  In this paper we show that combinations of levels
that are invariant under a global shift and a global rescaling gauge
configuration by gauge configuration, are,
perhaps surprisingly, in good agreement with those of the simplest RMM
imaginable.  Since the global shift and/or global rescaling can depend on
the gauge background they can have fluctuations of the same order of
magnitude as the energy levels themselves. To see whether this is true
we also look at eigenvalue observables that are invariant under
global rescalings and shifts only on the average. These observables
do not have a meaning gauge configuration by gauge configuration. 
We find that these observables are not described by the RMM.
We conclude that a
correct RMM might exist, but it would need to incorporate fluctuations
of the global scale and perhaps also of a global shift of the levels.
The difference from an ordinary RMM application is that the global
scale and shift variables cannot enter just as fixed parameters, but
need to be viewed as extra random variables.

While our data leaves little doubt that a simple RMM both works and
fails as indicated above, the amount of data that would be needed to
identify the correct RMM (assuming one exists) is beyond our
reach.

\section{Modelling the data.}

We wish to find a relationship between the distribution of the
$\lambda_j$'s and the $\xi_j$'s.
The Monte Carlo simulation produces sets of gauge links $C$, which, for the 
purposes of this discussion we take as totally independent samples from
the known probability distribution associated with the lattice
gauge theory action. The Dirac operator provides
one set of $\lambda_j (C),j=1,2,...$ for each gauge configuration. 
The distribution of the $C$'s induces a distribution of the sets of
$\lambda_j ,j=1,2,...$. We now view these sets of eigenvalues 
as new random variables. 

There is no doubt that a direct match
between the $\xi_j$'s and the $\lambda_j$'s is impossible:
$h_0$ is certainly non-universal, and obviously even the dimensions
of the $\lambda_j$ and of the $\xi_j$ do not match. 
One needs to allow at least for an extra scale and shift
when performing a comparison. If all that was needed in order
to obtain a match were a scale and a shift {\it parameter},
our study could easily proceed by looking at observables
made out of pairs of distinct eigenvalues. We focus on two examples,
defined below:
\be
c_{ij}=\frac{ \langle (\lambda_i - \langle \lambda_i \rangle )
(\lambda_j - \langle \lambda_j \rangle )\rangle }{\sqrt{ \langle
(\lambda_i - \langle \lambda_i \rangle )^2 \rangle \langle
(\lambda_j - \langle \lambda_j \rangle )^2\rangle }} 
\label{corr}
\ee

\be
r_{ij} = \frac{\sqrt{ <(\lambda_i - \lambda_j)^2> - <\lambda_i - \lambda_j>^2}}
{|<\lambda_i - \lambda_j>|}
\label{sigma}
\ee
These could be evaluated for the lattice data and compared
with the values obtained for the chRMM and the RMM. 
$c_{12}$ was referred to as $c$ in our previous work~\cite{prevplb}.

\TABLE{
\begin{tabular}{ccc}
$L$ & $N_c$ & $c_{12}$ \\
\hline
6 & 29 & 0.378(38) \\
6 & 37 & 0.343(49) \\
6 & 43 & 0.329(54) \\
7 & 17 & 0.416(50) \\
7 & 19 & 0.387(47) \\
7 & 23 & 0.410(43) \\
7 & 29 & 0.311(48) \\
8 & 13 & 0.343(51) \\
8 & 17 & 0.308(59) \\
8 & 23 & 0.310(59) \\
\hline
\end{tabular}
\caption{\label{tab1} The correlation $c_{12}$ at zero temperature. 
In the chRMM, $c_{12}=0.343(10)$.}}

To allow for fluctuations in the scale and shift variables, we
define a more restricted set of function,
associated with triplets of distinct eigenvalues. 
Again, these restricted variables are defined both in the
$\lambda_j (C)$ ensemble and in the $\xi_j$ ensemble.
For $ 1\le i < j < k $ let 
\be
r_{ijk}= \frac{\lambda_i-\lambda_j}{\lambda_i-\lambda_k}\label{ratio}
\ee
define ratios of differences. We define similar
quantities in the RMM and the chRMM and wish to compare their distributions.
The $r_{ijk}$'s are bounded:
\be
0\le r_{ijk} \le 1
\ee
If we imagine that for for every gauge configuration $C$,
there exist two hidden random variables, $a(C)$ and $b(C)$, it could be that
the variables $z_j (C)$, defined by
$z_j (C)=a(C)(\lambda_j (C) - b(C))$, \newline
are distributed the same way as the eigenvalues of $H$. 

To be sure, we do not know of a way to unambiguously 
determine whether the hidden variables $a(C)$ and $b(C)$ indeed exist, but
thinking in these terms leaves room for the case that they are 
just parameters, independent
of the gauge configuration $C$. Then, the correlations defined in eqn(\ref{corr}) and eqn(\ref{sigma}) and
the ratios defined in eqn(\ref{ratio}) for 
lattice data and the RMM would match. 
An analogous situation holds at low temperature, where 
$b(C)\equiv 0$ and $a(C)\equiv \Sigma N_c V$ ($V$ is the lattice volume)
determines the unrenormalized fermion condensate~\cite{nna}.
On the other hand, if $a(C)$ and $b(C)$
do depend upon $C$, the ratios defined in eqn(\ref{ratio}) would still match 
but the correlations in eqn(\ref{corr}) and eqn(\ref{sigma}) might not. 
Whatever the case may be, if the difference ratios match, we have established 
a well defined relationship between the Monte Carlo data and the 
RMM.\footnote{ To be sure, like everything else we do in this paper, 
there are the usual
caveats associated with numerical work that need to be kept in mind. 
This
fact will remain tacitly implied from now on.}

Our main point will end up being that the difference ratios match, but it is not
true that $a(C)$, $b(C)$ are just parameters. 
If $a(C)$ and $b(C)$ are at all meaningful
configuration by configuration our numerical work
shows that they fluctuate.

\subsection{Data analysis assuming a gauge field dependent scale and shift.}

\subsubsection{Statistics of eigenvalue pairs.}

We looked at $c_{12}$ in~\cite{prevplb} and concluded there that a
simple RMM does not work. This is not a consequence of some simulation
artefact: 
We start this section by showing that $c_{12}$ for lattice
data in the confined phase does match with $c_{12}$ for chRMM.
We use the $b=0.35$ lattice data in Table 1 of~\cite{nna}
for values of $L$ and $N_c$ where we found agreement with chRMM
using $<\lambda_1/\lambda_2>$. Since there is no shift, the
eigenvalue ratio is independent of scalings gauge configuration by
gauge configuration. It is only here that we check whether the
scale is indeed a parameter, independent of the gauge configuration. 
Table~\ref{tab1} in the present paper shows that the quantity
$c_{12}$ obtained
from the old data set also agrees with chRMM.~\footnote{All errors
quoted in this paper were obtained using the jackknife method.}
Thus, it is possible to confirm from the data that the eigenvalue
scale at low temperatures indeed is a parameter, as expected theoretically. 

\TABLE{
\begin{tabular}{ccccccc}
$ij$ & 
$c_{ij}^{\rm data}$ & $c_{ij}^{\rm RMM}$ & $c_{ij}^{\rm chRMM}$ &
$r_{ij}^{\rm data}$ & $r_{ij}^{\rm RMM}$ & $r_{ij}^{\rm chRMM}$\\
\hline
12 & 0.806(10) & 0.480(08) & 0.343(10) & 0.423(08) & 0.433(3) & 0.544(4)\\
13 & 0.678(16) & 0.346(09) & 0.188(10) & 0.291(06) & 0.262(2) & 0.339(2)\\
14 & 0.636(19) & 0.267(09) & 0.141(10) & 0.218(05) & 0.195(1) & 0.255(2)\\
15 & 0.585(29) & 0.234(09) & 0.107(10) & 0.179(06) & 0.156(1) & 0.200(1)\\
16 & 0.561(28) & 0.192(10) & 0.086(10) & 0.154(05) & 0.133(1) & 0.168(1)\\
23 & 0.799(11) & 0.553(07) & 0.447(08) & 0.462(09) & 0.429(3) & 0.471(3)\\
24 & 0.735(15) & 0.403(08) & 0.295(09) & 0.278(07) & 0.256(2) & 0.298(2)\\
25 & 0.674(23) & 0.344(09) & 0.211(09) & 0.208(07) & 0.184(1) & 0.220(2)\\
26 & 0.638(27) & 0.289(09) & 0.171(10) & 0.170(06) & 0.147(1) & 0.179(1)\\
34 & 0.846(09) & 0.581(07) & 0.511(08) & 0.424(10) & 0.431(3) & 0.452(3)\\
35 & 0.768(17) & 0.445(08) & 0.350(09) & 0.266(08) & 0.253(2) & 0.275(2)\\
36 & 0.715(20) & 0.365(09) & 0.268(09) & 0.201(06) & 0.184(1) & 0.206(2)\\
45 & 0.838(12) & 0.609(07) & 0.546(07) & 0.432(13) & 0.427(3) & 0.439(3)\\
46 & 0.770(16) & 0.468(08) & 0.395(09) & 0.260(08) & 0.252(2) & 0.266(2)\\
56 & 0.855(11) & 0.616(06) & 0.574(07) & 0.416(12) & 0.427(3) & 0.431(3)\\
\hline
\end{tabular}
\caption{\label{tab2} $c_{ij}$ and $r_{ij}$ in the deconfined phase compared
to the RMM and the chRMM.}}

In contradistinction to the confined phase, our previous work (last column
in Table 1 of~\cite{prevplb}) indicates that the $c_{12}$ extracted from lattice
data would disagree with any simple RMM. 
In this work we have carried out a more extensive study and confirmed this
finding. 
Table~\ref{tab2} shows
that the lattice data neither agrees with the RMM nor with the chRMM.
In addition, Table~\ref{tab2} also provides 
lattice results for the observables $r_{ij}$. The table displays
statistically significant differences from either the RMM or the chRMM
for several values of $(i,j)$, but the discrepancy is not dramatic. 
The weakness of the evidence might
be explained by bulk properties quickly dominating over edge features in 
the $r_{ij}$: the numbers from the RMM and the chRMM get close to each other
as $i$ departs from unity.

\subsubsection{Statistics of eigenvalue triplets.}

We look at averages and variances of the $r_{ijk}$. To get a
meaningful estimate for higher moments we would need substantially
more data. Table~\ref{tab3} shows agreement between lattice data and
our simple RMM, within errors, and the agreement
is quite meaningful. We also include in the table the prediction of
the chRMM
just to show that there is a measurable
difference. 

There is no question that there is a substantial 
numerical difference between
the cases of pairs and triplets of eigenvalues in the context
of random matrix modelling. 
The conclusions from Table~\ref{tab2} and Table~\ref{tab3}
about the deconfined phase are two fold: On the one hand there is evidence that 
$a(C)$ and $b(C)$ cannot be replaced by unfluctuating parameters.
On the other hand, for ratios of eigenvalue differences, 
a RMM with a soft spectral edge provides
substantial agreement with the lattice data.

\subsection{Data analysis assuming a matrix model with extra fluctuating variables.}

We have already learned that the data cannot be explained 
by setting a scale and a shift parameter to some (nonuniversal) values. 
To get a better handle on the effect 
we start afresh and define a hypothesis for a slightly 
extended class of RMM-s that still is compatible with the
assumption that ratios of eigenvalue differences in the data are distributed
identically to the same ratios in the simplest RMM. The basic change
in viewpoint is that we try to find an extension of the RMM
model that preserves the agreement for the eigenvalue-triplets but leaves
room to also explain the eigenvalue-pair properties. 

The hypothesis is presented below and what is meant by the double arrow 
is that the joint probability distribution of the variables
on the left hand side (indexed by $j$) 
is the same as the joint probability distribution of the variables
on the right hand side. 
\be
\lambda_j \leftrightarrow \alpha^\prime\xi_j+\beta^\prime
\ee
$\alpha^\prime$ and $\beta^\prime$ are random variables that 
have nonzero averages and relatively
small fluctuations around those averages. 
[With our conventions the average of $\alpha^\prime$ is negative.]

The probability distribution of the LHS variables is known in the sense 
that we know the lattice action and have an explicit expression for $A$ in
terms of the gauge configurations a simulation would produce. The probability
distribution of the variables on the RHS is not known. What we do
know (in the sense that it is part of the hypothesis)
about it is that, if we set $\alpha^\prime$ and $\beta^\prime$ to
fixed (and reasonable) values, the probability distribution of the $\xi_j$'s
is given by a standard RMM. Thus, much is known about the RHS yielding 
relations that the data would obey and thus providing tests we can carry
out. 

\TABLE
{\begin{tabular}{ccccccc}
\hline
$ijk$ & $r_{ijk}^{\rm data}$ & $r_{ijk}^{\rm RMM}$ & $r_{ijk}^{\rm chRMM}$
& $v_{ijk}^{\rm data}$ & $v_{ijk}^{\rm RMM}$ & $v_{ijk}^{\rm chRMM}$  \\
\hline
123 & 0.5519(43) & 0.5514(17) & 0.3520(18) & 0.0260(11) & 0.02901(47) & 0.02938(49) \\
124 & 0.3940(39) & 0.3951(13) & 0.1749(09) & 0.0166(08) & 0.01809(29) & 0.00856(17) \\
125 & 0.3085(44) & 0.3146(11) & 0.1044(05) & 0.0114(08) & 0.01243(20) & 0.00317(06) \\
126 & 0.2600(38) & 0.2647(10) & 0.0694(04) & 0.0086(06) & 0.00929(15) & 0.00141(03) \\
134 & 0.7198(35) & 0.7182(11) & 0.5124(15) & 0.0131(07) & 0.01298(22) & 0.02281(36) \\
135 & 0.5720(45) & 0.5707(10) & 0.3070(10) & 0.0119(08) & 0.01046(18) & 0.01003(18) \\
136 & 0.4820(42) & 0.4793(09) & 0.2044(07) & 0.0099(07) & 0.00824(14) & 0.00472(09) \\
145 & 0.7972(35) & 0.7965(08) & 0.6094(13) & 0.0070(05) & 0.00703(12) & 0.01743(28) \\
146 & 0.6707(34) & 0.6686(08) & 0.4068(10) & 0.0068(05) & 0.00643(11) & 0.00979(17) \\
156 & 0.8421(27) & 0.8407(07) & 0.6739(11) & 0.0041(03) & 0.00427(08) & 0.01299(21) \\
234 & 0.5336(53) & 0.5287(17) & 0.4092(17) & 0.0295(14) & 0.02934(47) & 0.02895(47) \\
235 & 0.3783(57) & 0.3691(13) & 0.2260(10) & 0.0187(13) & 0.01726(28) & 0.01051(19) \\
236 & 0.2980(47) & 0.2884(11) & 0.1449(07) & 0.0129(09) & 0.01130(19) & 0.00455(08) \\
245 & 0.7057(48) & 0.7007(12) & 0.5644(14) & 0.0135(09) & 0.01391(24) & 0.02057(33) \\
246 & 0.5536(43) & 0.5468(10) & 0.3626(10) & 0.0107(07) & 0.01067(18) & 0.01058(18) \\
256 & 0.7858(35) & 0.7824(09) & 0.6499(12) & 0.0071(05) & 0.00753(13) & 0.01463(24) \\
345 & 0.5221(72) & 0.5209(17) & 0.4368(17) & 0.0296(20) & 0.02906(45) & 0.02923(47) \\
346 & 0.3597(53) & 0.3590(13) & 0.2540(11) & 0.0164(11) & 0.01682(27) & 0.01191(21) \\
356 & 0.6929(47) & 0.6922(12) & 0.5909(14) & 0.0129(08) & 0.01414(24) & 0.01879(30) \\
456 & 0.5140(70) & 0.5154(17) & 0.4504(17) & 0.0284(19) & 0.02874(46) & 0.02867(46) \\
\hline
\end{tabular}
\caption{\label{tab3}
Lattice data for $r_{ijk}$ and $v_{ijk}$ compared with the RMM and the chRMM.}
}

We set our test up by defining $\mu_j=\lambda_j-\lambda_1,~j=2,3..$ and
$\xi_1-\xi_j=\eta_j,~j=2,3..$. According to the hypothesis:
\be
\mu_j \leftrightarrow -\alpha^\prime\eta_j
\ee
In terms of the variables $\ln \mu_j$ and $\ln \eta_j$ we have 
\be
\ln \mu_j \leftrightarrow \ln \eta_j+\ln(-\alpha^\prime)
\ee
This produces the following relation, for all $j\ge 2$:
\be
<\ln \mu_j > - <\ln \eta_j > = <\ln (-\alpha^\prime)>
\label{eqn1}
\ee
The LHS in the above equation can be evaluated: the first term from the
data and the second from what we know about the extended RMM. We get $j=2,3...$ determinations
of the right hand side which must agree with each other.
The results from the lattice data checking the independence
of the RHS of equation (\ref{eqn1}) on $j$ are shown in Table~\ref{tab4}.
In conclusion, our hypothesis has survived a test and has produced a number
for the average of one of the new random variables (more precisely, of its logarithm).

\TABLE{
\begin{tabular}{cccc}
$N_c$ & $j$ & $<\ln \mu_j>$ & $<\ln (-\alpha^\prime)>$ \\
\hline
53 & 2 & -5.376(26) & -2.855(26)    \\
53 & 3 & -4.732(16) & -2.866(16)    \\
47 & 2 & -5.325(13) & -2.803(14)    \\
47 & 3 & -4.679(08) & -2.813(08)    \\
47 & 4 & -4.335(07) & -2.814(07)    \\
47 & 5 & -4.103(07) & -2.816(08)    \\
47 & 6 & -3.929(06) & -2.818(06)    \\
43 & 2 & -5.259(28) & -2.737(28)    \\
43 & 3 & -4.607(15) & -2.741(15)    \\
37 & 2 & -5.171(28) & -2.649(28)    \\
37 & 3 & -4.491(15) & -2.625(15)    \\
\hline
\end{tabular}
\caption{\label{tab4} 
LHS of eqn(\ref{eqn1}) versus $j$.}
}

We now proceed to look at fluctuations. Define:
\begin{eqnarray}
\label{fluct}
\Delta_j &=& \ln \mu_j - <\ln \mu_j >,~
\delta_j =\ln\eta_j - <\ln \eta_j >, \cr
 \delta &=& \ln(-\alpha^\prime ) -
<\ln (-\alpha^\prime ) >
\end{eqnarray}
Using the above definitions we get:
\begin{eqnarray}
\label{eqn2}
<\Delta_j \Delta_k > - < \delta_j \delta_k > = \cr
<\delta^2> +<\delta(\delta_j + \delta_k)>
\end{eqnarray}
The first term on the LHS is obtained from the data and the second
term on the LHS from the simplest, unextended, RMM. There is little we
know about the RHS, except that the dependence on the indices $j$ and
$k$ is through quantities that enter linearly.  The results are shown
in Table~\ref{tab5}.  

If $\alpha^\prime$ were a fixed parameter,
$\delta\equiv 0$ and the RHS of equation (\ref{eqn2}) would be
zero. Except for $(j,k)$ equal to $(2,3)$, $(2,4)$, $(2,5)$
and $(2,6)$ all other entries indicate that the RHS is not zero.
We therefore admit that
the fluctuation $\delta^2$ cannot be neglected.  

A simple possibility
would be that there are no correlation between $\delta$ and
$\delta_j$, in which case the LHS would need to emerge positive and
independent of $jk$. For $(j,k)$ equal to $(2,2)$ the entry in the table 
is negative
and significantly away from zero. Furthermore, the non-zero
LHS entries are not all equal indicating that 
there are correlations between $\delta$ and
$\delta_j$. 
In principle, it would be possible to extract $<\delta^2$ and
$<\delta\delta_j>$ using the lattice data, but the sample of 
eigenvalue sets $\{ \lambda_j \}$ is too small for this. 

Eqn(\ref{eqn2}) can be used to eliminate the dependence on $\delta$, providing
a test that a fluctuating $\alpha^\prime$ might be a correct way to describe
the data. 
\be
\label{fluct3}
<\Delta_j \Delta_k > -<\delta_j \delta_k > =
\frac{<\Delta_j^2> - <\delta_j^2>}{2} + \frac{<\Delta_k^2>-<\delta_k^2>}{2}
\ee
The data indeed is 
consistent with the above relation, but 
much better accuracy is needed to make this test convincing.

The above analysis, and the other entries in the table show that substantially
higher accuracy would be needed to convince one that the extended model is correct.
Achieving this accuracy is beyond our numerical capacity. We only see 
that a modest extension of the simplest RMM could
provide a description of the data and that the simplest
RMM is unlikely to work, in agreement with the previous section. 

This concludes our analysis of the Dirac eigenvalues data we generated. 
In the interest of brevity we have not presented data at smaller values of $N_c$,
where the agreement of triplet eigenvalue observables with the RMM has not yet set in.

\TABLE{
\begin{tabular}{cccc}
$jk$ & $<\Delta_j\Delta_k>$ & $<\delta_j\delta_k>$ &
$<\Delta_j\Delta_k> - <\delta_j\delta_k>$  \\
\hline
22 & 0.2201(102) & 0.2509(51) & -0.0308(114) \\
23 & 0.0959(46)  & 0.0916(18) &  0.0043(50)  \\
24 & 0.0624(37)  & 0.0608(12) &  0.0016(39)  \\
25 & 0.0480(41)  & 0.0469(09) &  0.0011(42)  \\
26 & 0.0403(35)  & 0.0391(08) &  0.0012(36)  \\
33 & 0.0902(37)  & 0.0751(12) &  0.0151(39)  \\
34 & 0.0539(26)  & 0.0429(07) &  0.0110(27)  \\
35 & 0.0423(29)  & 0.0317(05) &  0.0106(29)  \\
36 & 0.0346(24)  & 0.0261(05) &  0.0085(24)  \\
44 & 0.0470(21)  & 0.0397(06) &  0.0074(22)  \\
45 & 0.0338(22)  & 0.0262(04) &  0.0076(23)  \\
46 & 0.0276(19)  & 0.0210(03) &  0.0066(19)  \\
55 & 0.0314(19)  & 0.0248(04) &  0.0066(20)  \\
56 & 0.0242(16)  & 0.0181(03) &  0.0061(16)  \\
66 & 0.0231(14)  & 0.0178(03) &  0.0053(14)  \\
\hline
\end{tabular}
\caption{\label{tab5} 
Evidence for fluctuations in the scale from (\ref{eqn2}). }
}

\section{Physical relevance.}

How could random matrix theory be useful to understand the planar limit ? 
The answer is:
if there is a RMM that applies, one knows how the large $N_c$ limit is approached.
This is very useful if the approach is controlled by a physically 
relevant parameter, as at low temperature, where the parameter
is the bi-fermion condensate. It can also be useful
when one wants to establish by numerical means that a conjectured property of 
the large $N_c$ limit indeed holds. 

Another physical observable that a RMM can help quantify is
the behavior of the spectral density, $\rho(\lambda)$ close
to the gap, $\lambda_g$.
If, on the finite lattice in the deconfined phase, 
$\rho(\lambda)$ is a smooth function close to $\lambda_g$, 
because of large $N_c$ reduction, this function is the same 
on an infinite lattice. On an infinite lattice, $\rho(\lambda)$ would be a smooth
curve even at finite $N_c$ and may even be a good approximation
to the spectral curve in full QCD. 
The infinite $N_c$ curve has an end point where the 
gap starts and some specific structure at that endpoint.

We have argued in ~\cite{prevplb}
that the gap not only exists on the lattice, 
but also has a reasonable continuum limit when
$N_c=\infty$. 
We would like to eventually be able to make a similar statement
about the structure at the spectral end point. If a simple RMM
applies we expect the continuous eigenvalue density $\rho(\lambda)$
to go as:
\be
\rho(\lambda)\propto (\lambda-\lambda_g)^{1/2}~~{\rm~random~matrix~prediction}
\ee
where $\lambda_g$ is the gap energy.
On the other hand, for high temperatures one might expect a perturbative
spectrum, with $\lambda_g \propto \pi T$. The density of states  for
$\lambda\approx\lambda_g$ would go as
\be
\rho(\lambda)\propto (\lambda-\lambda_g)^{(d-3)/2}~~{\rm~free~field~perturbation~theory~prediction}
\ee
in $d$ dimensional Euclidean space.
While the RMM formula comes from a framework that is oblivious of the
dimensionality of the system, the perturbative formula matches
this only because spacetime is four dimensional. The degree to which this
accident is responsible for our numerical findings is worthy of further
study. 

It is unknown whether perturbation theory makes the correct prediction
for the behavior of the spectrum of the Dirac operator at the edge; if we determined that
a particular RMM is supported by Monte Carlo
data and also established the square root behavior at the spectral
edge, we would have learned something. 
The square root behavior requires a new dimensionful parameter as its coefficient: 
Using the right RMM, one might be able to numerically determine this parameter, in addition
to the gap energy. 

The structure of the eigenvalue density of the massless Dirac operator in the deconfined
phase is related to current--current correlations. We leave to further study
what implications this might have.

\section{Conclusions.}

We saw some evidence that fluctuations 
of the eigenvalues of the Dirac operator in the deconfined regime 
behave differently from fluctuations in simple random matrix models.
We also presented quite substantial evidence
that there is much in common between the statistics of the spectrum 
of the Dirac operator and that of a simple matrix model. 
It is possible to reconcile these two trends in an extended matrix model,
but our data is too meager to convincingly establish the validity of
such an extended model.
If this could be done, one might learn something about the way the
spectral density vanishes at the edge of its support.
If this effect were to be shown to extend to the continuum limit we
might learn something about current--current correlations in the QCD plasma
in the planar approximation.

\acknowledgments

R. N. acknowledges support by the NSF under grant number
PHY-055375 and from Jefferson Lab. The Thomas Jefferson National
Accelerator Facility (Jefferson Lab) is operated by the
Southeastern Universities Research Association (SURA) under DOE
contract DE-AC05-84ER40150. H. N. acknowledges partial
support by the DOE under grant number DE-FG02-01ER41165 at Rutgers,
an Alexander von Humboldt award and the hospitality of the Physics
department at Humboldt University, Berlin. 
We thank Poul Damgaard and Jac Verbaarschot for useful discussions.

\end{document}